From Bernard.Helffer@math.u-psud.fr  Tue Feb  8 10:23:40 2000
Received: from nef.ens.fr (nef [129.199.96.32]) by clipper.ens.fr (8.9.2/jb-1.1)
	id KAA13970 for <dutour@clipper.ens.fr>; Tue, 8 Feb 2000 10:23:40 +0100 (MET)
Return-Path: <Bernard.Helffer@math.u-psud.fr>
Received: from matups.math.u-psud.fr (matups.matups.fr [129.175.50.4])
          by nef.ens.fr (8.9.3/1.01.28121999) with ESMTP id KAA00952
          for <dutour@clipper.ens.fr>; Tue, 8 Feb 2000 10:23:39 +0100 (CET)
Received: from lanors.math.u-psud.fr (lanors.math.u-psud.fr [129.175.53.10])
          by matups.math.u-psud.fr (8.9.1a/jtpda-5.3.1) with ESMTP id KAA13350
          for <dutour@clipper.ens.fr>; Tue, 8 Feb 2000 10:23:38 +0100 (MET)
Received: from lan62.math.u-psud.fr (lan62 [129.175.53.62])
	by lanors.math.u-psud.fr (8.9.3/8.9.3) with ESMTP id KAA18656
	for <dutour@clipper.ens.fr>; Tue, 8 Feb 2000 10:23:36 +0100 (MET)
Received: (from helffer@localhost)
	by lan62.math.u-psud.fr (8.9.3+Sun/8.9.1) id KAA15571
	for dutour@clipper.ens.fr; Tue, 8 Feb 2000 10:23:35 +0100 (MET)
Date: Tue, 8 Feb 2000 10:23:35 +0100 (MET)
From: Bernard Helffer <Bernard.Helffer@math.u-psud.fr>
Message-Id: <200002080923.KAA15571@lan62.math.u-psud.fr>
To: dutour@clipper.ens.fr
Status: RO

\documentclass{article}
\def\nz{\ifmmode {I\hskip -3pt N} \else {\hbox {$I\hskip -3pt N$}}\fi}
\def\zz{\ifmmode {Z\hskip -4.8pt Z} \else
       {\hbox {$Z\hskip -4.8pt Z$}}\fi}
\def\qz{\ifmmode {Q\hskip -5.0pt\vrule height6.0pt depth 0pt
       \hskip 6pt} \'else {\hbox
       {$Q\hskip -5.0pt\vrule height6.0pt depth 0pt\hskip 6pt$}}\fi}
\def\rz{\ifmmode {I\hskip -3pt R} \else {\hbox {$I\hskip -3pt R$}}\fi}
\def\cz{\ifmmode {C\hskip -4.8pt\vrule height5.8pt\hskip 6.3pt} \else
       {\hbox {$C\hskip -4.8pt\vrule height5.8pt\hskip 6.3pt$}}\fi}

\def\const{{\rm const.\,}} 
\def\div{{\rm div}\,}
\def\rot{{\rm rot}\,} 
\def\rotadj{{\rm rot^* }} 
\def\supp{\mathop{\rm supp} \nolimits} 

\def\qed{\hbox {\hskip 1pt \vrule width 4pt height 6pt depth 1.5pt
        \hskip 1pt}\\}
\def\Fg {{\cal F}} 
\def\Eg {{\cal E}} 
\def\Lg {{\cal L}} 
\def\Og {{\cal O}} 
\def\Rg {{\cal R}} 
\def\Sg {{\cal S}} 
\def\and {{\rm \; and \;}}

\def\Im {{\rm \; Im\;}}
\def\Re {{\rm \;Re\;}}
\newcommand {\pa}{\partial}
\newcommand {\ar}{\rightarrow}

\makeatletter
\@addtoreset{equation}{section}

\makeatother

\newtheorem{theorem}{Theorem}[section]
\newtheorem{lemma}[theorem]{Lemma}
\newtheorem{proposition}[theorem]{Proposition}
\newtheorem{definition}[theorem]{Definition}
\newtheorem{remark}[theorem]{Remark}
\newtheorem{corollary}[theorem]{Corollary}

\title{On bifurcations from normal solutions for superconducting states}
\author{Mathieu Dutour, Bernard Helffer\\
UMR 8628 CNRS,
\\
D\'epartement de Math\'ematiques,\\
Universit\'e Paris-Sud, F-91405 Orsay.
}
\date{February  07, 2000} 
        
\begin{document}

\bibliographystyle{plain}
\maketitle
%
%
\begin{abstract}
Motivated by the paper by J.~Berger and K.~Rubinstein \cite{BeRu} and
 other recent studies \cite{GiPh}, \cite{LuPa1}, \cite{LuPa2}, 
 we analyze the Ginzburg-Landau functional in an open bounded set $\Omega$. We mainly discuss the bifurcation problem
whose analysis was initiated in \cite{Od} and show how some of 
the techniques developed by the first author  
 in the case of Abrikosov's superconductors \cite{Du} can be applied in this context. In the case
 of non simply connected domains, we come back to \cite{BeRu} and \cite{HHOO}, \cite{HHOO1}
 for giving the analysis of the structure of the nodal sets for the bifurcating solutions. \end{abstract} 
\section{Introduction}\label{Section1n}
\subsection{Our model.}\label{Subsection1.1}
Following the paper by Berger-Rubinstein \cite{BeRu}, we would like to understand the
 minima (or more generally the extrema) of the following Ginzburg-Landau functional.
 In a bounded, connected, regular\footnote{with
 $C^\infty$ boundary}, open set $\Omega\subset \rz^2$ and, for any $\lambda >0$ and $\kappa >0$,
this functional  $G_{\lambda,\kappa}$ is defined,
 for $u\in H^1(\Omega;\cz)$ and $A\in H_{loc}^1(\rz^2; \rz^2)$ such that $\rot A \in L^2$, by
\begin{equation}\label{i1}
\begin{array}{ll}
G_{\lambda,\kappa} (u,A)& = 
\int_\Omega \left(\lambda ( - |u|^2 + \frac 12 |u|^4) + |(\nabla - iA) u|^2\right)) dx_1\cdot dx_2\\
 &\quad + \kappa^2 \lambda^{-1} \int_{\rz^2} |\rot A - H_e|^2 dx_1\cdot dx_2\;.
\end{array}
\end{equation}
Here, for $A=(A_1,A_2)$, $\rot A = \pa_{x_1} A_2 - \pa_{x_2} A_1$,  $\div A = \pa_{x_1} A_1
 + \pa_{x_2} A_2$
 and $H_e$ is a $C_0^\infty$ function on $\rz^2$ (or more generally some function
 in $L^2(\rz^2)$).
Physically $H_e$ represents the exterior magnetic field.\\
Let $A_e$ be a solution of 
\begin{equation}\label{i2}
\begin{array}{l}
\rot A_e = H_e \\
\div A_e = 0\;.
\end{array}
\end{equation}
It is easy to verify that such a solution exists by looking for $A_e$ in the form
 $A_e= ( - \pa_{x_2} \psi_e, \pa_{x_1} \psi_e)$. We have then to solve $\Delta \psi_e = H_e$
 and it is known to be solvable in $\Sg'(\rz^2) \cap C^\infty (\rz^2)$ (or in $\Sg'(\rz^2) \cap H^2_{loc}(\rz^2)$
 if $H_e\in L^2(\rz^2)$). Of course $A_e$ is not
 unique but we shall discuss about uniqueness modulo gauge transform later and at the end
 this is mainly
 the restriction of $A_e$ to $\Omega$ which will be considered.\\
We shall sometimes use the identification between vector fields $A$ and $1$-forms $\omega_A$.\\

When analyzing the extrema of the GL-functional, it is natural to first analyze
 the corresponding Euler-Lagrange equations.
This is a system of two equations (with a boundary equation)~:
\begin{equation}\label{i3}
\begin{array}{ll}
(GL)_1 &- (\nabla - i A)^2 u + \lambda u ( |u|^2 -1) = 0 \;,\; \mbox{ in } \Omega\;,\\
(GL)_2 & \rotadj (\rot A - H_e) = \lambda \kappa^{-2} \Im \left[ {\bar u}\cdot 
(\nabla - i A) u \right]\cdot 1_\Omega\;,\\
(GL)_3 & (\nabla -i A) u\cdot \nu = 0 \;,\; \mbox{ in } \pa \Omega\;.
\end{array}
\end{equation}
Here $\nu$ is a unit exterior normal to $\pa \Omega$. The operator $\rotadj$ is defined by
$\rotadj f:= (\pa_{x_2} f, - \pa_{x_1} f)$.\\
Moreover, without loss of generality in our problem, we shall add the condition
\begin{equation}\label{i4}
(GL)_4 \qquad \div A = 0\; \mbox{ in } \Omega.
\end{equation}
One can also assume if necessary that
 the vector potential satisfies
\begin{equation}\label{a18}
A\cdot \vec{\nu} = 0\;,
\end{equation}
on the boundary of $\Omega$, where $\nu$ is a normal unit vector to $\pa \Omega$.\\
Let us briefly recall the
 argument. One would like to find $\theta$ in $C^\infty({\bar
 \Omega})$ such that ${\tilde A} = A + d\theta$ satisfies (\ref{i4})
 and (\ref{a18}). One can proceed in two steps. The first step is to
 find
 a gauge transformation such that (\ref{a18}) is satisfied. This is
 immediate if the boundary is regular.\\ We now
 assume this condition. \\
The second step  consists in solving 
$$
\begin{array}{l}
\Delta \theta = - \div A\;\mbox{ in } \Omega, \\
\frac{\pa \theta}{\pa \nu}= 0,\mbox{ on } \pa \Omega\;.
\end{array}
$$
This is a Neumann problem,  which is solvable iff the right
hand-side is orthogonal to the first eigenfunction of the Neumann
realization of
 the Laplacian, that is the constant function $x \mapsto 1$. We have only to observe that
$\int_\Omega \div A \; dx = 0$ if (\ref{a18}) is satisfied.\\

An important remark is that the pair $(0,A_e)$ is a solution of the system. This solution is called
 the normal solution. Of course, any solution of the form
 $(0, A_e + \nabla \phi)$ with $\phi$ harmonic is also a solution.\\
\begin{remark}\label{Remark1.1}.\\
{\rm Note also that the normalization of the functional leads to the property that
\begin{equation}\label{a1}
G_{\lambda, \kappa} (0, A_e)=0\;.
\end{equation}
}
\end{remark}
The first proposition is standard.
\begin{proposition}\label{Proposition1.2}.\\
If $\Omega$ is bounded, the functional $G_{\lambda,\kappa}$ admits a global minimizer which is 
a solution of the equation.
\end{proposition}
We refer to \cite{DGP}, for a proof together with the discussion of
the next subsection.
\subsection{Comparison with other models}\label{Subsection1.2}
Let us observe that there is another natural problem which may be considered. This is the problem
 of minimizing, for $(u,A)\in H^1(\Omega,\cz)\times H^1(\Omega, \rz^2)$, the functional $G_\lambda^\Omega$
 defined by
\begin{equation}\label{i6}
\begin{array}{ll}
G_{\lambda,\kappa}^\Omega (u,A)& = 
\int_\Omega \left(\lambda ( - |u|^2 + \frac 12 |u|^4) + |(\nabla - iA) u|^2\right)\; dx_1\cdot dx_2\\
 &\quad + \kappa^2 \lambda^{-1} \int_{\Omega} |\rot A - H_e|^2 \; dx_1\cdot dx_2\;.
\end{array}
\end{equation}
This may  lead to a different result in the case when $\Omega$ is not simply connected.
According to discussion with Akkermans, this is the first problem which is the most
 physical (See also the discussion in the appendix).\\
A comparison between $G_{\lambda,\kappa}$ and  $G_{\lambda, \kappa}^{\Omega,D}$ where $D$ is a  ball
 containing $\Omega$ and $G_{\lambda,\kappa}^{\Omega,D}$ is defined by
\begin{equation}\label{i7}
\begin{array}{ll}
G_{\lambda,\kappa}^{\Omega, D} (u,A)& = 
\int_\Omega \left(\lambda ( - |u|^2 + \frac 12 |u|^4) + |(\nabla - iA) u|^2\right) \; dx_1\cdot dx_2\\
 &\quad + \kappa^2 \lambda^{-1} \int_{D} |\rot A - H_e|^2 \; dx_1\cdot dx_2\;.
\end{array}
\end{equation}
is useful.  If  $b$  is given with support outside of the ball $D$, it
 is 
 easy to see (assuming that $b$ is regular) that there exists $a$  with support
 outside $D$ such that $\rot a = b$. It is indeed sufficient to take the usual transversal
 gauge 
\begin{equation}\label{i8}
a_1 =- x_2 \int_0^1 s b(sx) ds\;,\; a_2 = x_1 \int_0^1 s b(sx) ds\;.
\end{equation}
This shows that, for any $D$ containing $\Omega$,
we have
\begin{equation}\label{i9}
\inf G_{\lambda,\kappa} (u,A) = \inf  G_{\lambda,\kappa}^{\Omega,D} (u,A) \;.
\end{equation}
In particular it is enough to consider minimizing sequences $(u_n, A_e + a_n)$ where
 $\supp a_n \subset D$ and $D$ is a ball containing $\Omega$. The proof of the existence of minimizers
 is then greatly simplified.\\
Finally, it is natural \footnote{This is at least clear when $\tilde \Omega$
 is a  star-shaped domain by the previous proof. See Section
 \ref{Section3n}, in the proof of Proposition \ref{Proposition3.2} for a complementary argument.} to think that one can replace $D$ by 
\begin{equation}\label{i9a}
{\tilde \Omega}:=\Omega\cup_i \Og_i\;,
\end{equation}
 where the $\Og_i$ are the holes, that are the bounded connected components
 of $\rz^2\setminus \Omega$. A proof can be obtained by analyzing the Ginzburg-Landau equations satisfied
 by a minimizer of $G_{\lambda,\kappa}^{\Omega, D}$. We finally get~:
\begin{equation}\label{i9b}
\inf G_{\lambda,\kappa} (u,A) = \inf  G_{\lambda,\kappa}^{\Omega,{\tilde \Omega}} (u,A) \;.
\end{equation}

\begin{remark}\label{Remark1.3}.\\{\rm 
If $(u,A_e+a)$ is a solution of the GL-equation then $\rot a =0$ in the unbounded
 component of $\rz^2\setminus \Omega$ and $\rot a = \const$ in each hole (See Lemma~2.1
 in \cite{GiPh}).  It would be
 interesting to discuss what the possible values of these constants are~: are they equal 
to $0$~?}
\end{remark}
\subsection{Standard results}\label{Subsection1.3}

The second proposition which is also quite standard (See for example
\cite{DGP}) is
\begin{proposition}\label{Proposition1.5}.\\
If $u$ is a solution  of the first GL-equation with the Neumann boundary condition
 then
\begin{equation}\label{a2}
|u(x)|\leq 1\;,\; \forall x\in \Omega\;.
\end{equation}
\end{proposition}
We note for further use that the solutions of the G-L system are in $C^\infty (\overline{\Omega})$
 under the assumption that $\Omega$ is regular.

\section{ Is the normal state a minimizer~? \label{Section2n}}
The aim of this section is to give a proof of a result suggested in \cite{BeRu}
 who said `` We expect the normal state to be a stable solution
 for small $\lambda$...''.\\ Although, this result is probably known as folk theorem,
 we think it is useful to give a proof (following considerations by M.~Dutour
 in a near context \cite{Du}) of this property.\\
Note that connected results are obtained   in \cite{GiPh} and more
 recently in \cite{LuPa1}, \cite{LuPa2}.\\
Before  stating  the theorem, let us recall that we have called {\it normal state} a pair
 $(u, A)$ of the form~:
\begin{equation}\label{2.1}
(u,A) = (0, A_e)\;,
\end{equation}
where $A_e$ is any solution of (\ref {i1}).\\
We note that this is well defined up to gauge transformation. Moreover, we have~:
 \begin{lemma}\label{Lemma2.1}.\\
$(0,A_e)$ is a solution of the GL-system.
\end{lemma}
So it is effectively natural to ask if $(0,A_e)$ is a global minimum.
The first result in this direction is
 the following easy proposition about the normal state. But let us first
 introduce~:
\begin{definition}\label{Definition2.2}.\\
We denote by $\lambda^{(1)}$ the lowest eigenvalue of the Neumann realization in $\Omega$
of 
$$ - \Delta_{A_e}:= - (\nabla - i A_e)^2\;.
$$
\end{definition}
We shall frequently use the assumption
\begin{equation}\label{x0}
\lambda^{(1)} >0\;.
\end{equation}
Note the following necessary and sufficient condition for this property (cf \cite{He}).
\begin{proposition}\label{BoAh}.\\
The condition (\ref{x0}) is satisfied if and only if 
 one of two following conditions is satisfied~:
\begin{enumerate}
\item   $H_e$ is not identically zero in $\Omega$;
\item  $H_e$ is  identically zero in $\Omega$ but there exists a closed path $\gamma$ 
in $\Omega$
 such that $\frac{1}{2\pi} \int_\gamma \omega_{A_e} \not\in \zz$.
\end{enumerate}
\end{proposition}
Let us observe that the second case can only occur when $\Omega$ is
 non simply connected.

\begin{proposition}\label{Proposition2.7}.\\
Under condition (\ref{x0}) and if $\lambda\in]0,\lambda^{(1)}[$, the pair
$(0,A_e)$ is  a non-degenerate (up to gauge transforms) local minimum  of $G_{\lambda,\kappa}$.
\end{proposition}
The Hessian at $(0,A_e)$ of the GL-functional is indeed the map
$$
(\delta u,\delta a) \mapsto ((-\Delta_{A_e}-\lambda)\, \delta u\,,\, \rotadj\, \rot \delta a)\;,
$$
where we assume that $\div \; \delta a =0$ and $\delta a \cdot \nu =0$ at the boundary
 of $\tilde \Omega$.\\
Note that this proof gives also~:
\begin{proposition}\label{Proposition2.7a}.\\
If $\lambda >\lambda^{(1)}$, the pair
$(0,A_e)$ is not  a local minimum  of $G_{\lambda,\kappa}$.
\end{proposition}
We refer to \cite{LuPa1} for a connected result.
Proposition \ref{Proposition2.7a} does not answer completely to the
question about global minimizers. The next theorem gives a
complementary information.

\begin{theorem}\label{Theorem2.2}.\\
Under assumption (\ref{x0}), then, for any $\kappa >0$,  there exists $\lambda_0(\kappa) >0$ such that, for
 $\lambda\in ]0,\lambda_0 (\kappa)]$,
 $G_{\lambda,\kappa}$ has only normal solutions as global minimizers.
\end{theorem}
\begin{remark}\label{Remark2.4}.\\
{\rm Using a variant of the techniques used in \cite{Du} in a similar context, one can actually show that,
  for any $\kappa>0$,  there exists $\lambda_1(\kappa) >0$ such that, for
 $\lambda\in ]0,\lambda_1 (\kappa)]$,
 $G_{\lambda,\kappa}$ has only the normal solutions
 as solutions of the Ginzburg-Landau equations. This will be analyzed
  in Section \ref{Section5n}.}
\end{remark}
{\bf Proof of Theorem \ref{Theorem2.2}}:\\
Let $(u,A):=(u, A_e + a)$ be a minimizer of the $(GL)$ functional. So it is a solution\footnote{
 We actually  do not use
 this property in the proof.}
 of (GL) and moreover we have, using (\ref{a1}), the following property~:
\begin{equation}\label{x1}
G_{\lambda,\kappa} (u, A)\leq 0\;.
\end{equation}
Using the inequality $-|u|^2 \geq -\frac 12 |u|^4 - \frac 12$ and (\ref{x1}), we first get, with $b=\rot a$~:
\begin{equation}\label{x2}
\frac{\kappa^2}{\lambda} \int_{\rz^2} b^2 \; dx \leq \frac \lambda 2 |\Omega|\;,
\end{equation}
where $|\Omega|$ is the area of $\Omega$.

We now discuss the link between $b$ and $a$  in ${\widetilde \Omega}$. So we shall only use
 from (\ref{x2})~:
\begin{equation}\label{x3}
\frac{\kappa^2}{\lambda} \int_{\widetilde \Omega} b^2 \; dx \leq \frac \lambda 2 |\Omega|\;,
\end{equation}
Let us now consider in ${\widetilde \Omega}$, ${\tilde a}$ the problem of finding
 a solution of
\begin{equation}\label{x4}
\begin{array}{ll}
\rot \;{\tilde a} = b\;,\;& \div{\tilde a} = 0\;,\; \\
{\tilde a}\cdot \nu= 0\;,\; \mbox{ on } \pa{\widetilde \Omega}\;.
\end{array}
\end{equation}
We have the following standard proposition (see Lemma~2.3 in \cite{GiPh}).
\begin{proposition}\label{Proposition2.3}.\\
The problem (\ref{x4}) admits, for any $b\in L^2({\widetilde \Omega})$, a unique solution
 $\tilde a$ in $H^1({\widetilde \Omega})$. Moreover, there exists a constant $C$ such that
\begin{equation}\label{x4a}
||{\tilde a}||_{H^1({\widetilde \Omega})} \leq C\, || b||_{L^2({\widetilde \Omega})}\;,\; \forall b\in L^2\;.
\end{equation}
\end{proposition}
{\bf Proof of Proposition \ref{Proposition2.3}.}\\
Following a suggestion of F.~Bethuel, we look for a solution in the form~: 
${\tilde a} = \rot^*\psi$. We then solve the Dirichlet problem $-\Delta \psi = b$
 in $\widetilde \Omega$. This gives a solution with the right regularity. For the uniqueness,
 we observe that $\widetilde \Omega$ being connected and simply connected a solution of
 $\rot {\hat a} =0$ is of the form $\hat a = d \theta$
 (with $\theta\in H^{2}({\widetilde \Omega})$, 
 and if $\div {\hat a} =0$
 and ${\hat a}\cdot \nu$ on $\pa {\widetilde \Omega}$, we get
 the equations $\Delta \theta =0$
 and $\nabla \theta\cdot \nu =0$ on $\pa {\widetilde \Omega}$, which implies $\theta=\const$
 and consequently $\hat a=0$.\qed\\
We can now use the Sobolev estimates in order to get
\begin{equation}\label{x5}
||a||_{L^4 ({\widetilde \Omega})} \leq C_1\; || a ||_{H^1({\widetilde \Omega})}\;.
\end{equation}
>From (\ref{x3}), (\ref{x4a}) and (\ref{x5}), we get the existence of a constant $C_2$ such 
 that
\begin{equation}\label{x5a}
 ||a||_{L^4({\widetilde \Omega})} \leq C_2 \;\frac{\lambda}{\kappa}\;.
\end{equation}

The second point is to observe, that, for any $\epsilon \in]0,1[$, we have the inequality
\begin{equation}\label{x6}
\int_\Omega | (\nabla - iA) u|^2 \; dx \geq (1-\epsilon)
 ||(\nabla -i A_e) u ||_{L^2(\Omega)}^2 - \frac{(1-\epsilon)}{\epsilon} || a u||_{L^2(\Omega)}^2 \;.
\end{equation}
Taking $\epsilon = \frac 14$ and  using H\"older's inequality, we  get
\begin{equation}\label{x6a}
\int_\Omega | (\nabla - iA) u|^2\; dx \geq \frac 34
 ||(\nabla -i A_e) u ||_{L^2(\Omega)}^2 - 3 || a ||_{L^4(\Omega)}^2
 || u||_{L^4(\Omega)}^2\;.
\end{equation}
Using now the ellipticity of $-\Delta_{A_e}$ in the form of the existence of a constant $C_1$
\begin{equation}\label{x6b}
||u ||^2_{H^1(\Omega)} \leq C_1 \; \left( || (\nabla - i A_e) u ||_{L^2(\Omega)}^2
  + ||u||_{L^2(\Omega)}^2 \right)\;,
\end{equation}
and again the Sobolev inequality,
 we then obtain the existence of a constant $C_2$ such that
\begin{equation}\label{x6a1}
\int_\Omega | (\nabla - iA) u|^2 \;dx \geq \left(\frac 34  - C_2 ||a||_{L^4(\Omega)}^2 \right) 
||(\nabla -i A_e) u ||_{L^2(\Omega)}^2 - C_2
|| a ||_{L^4(\Omega)}^2
 || u||_{L^2(\Omega)}^2\;.
\end{equation}
 We get then from (\ref{x1}) and (\ref{x5a}), and for a suitable new constant  $C$ 
(depending only on $\Omega$ and $H_e$),
\begin{equation}\label{x7}
\left[ \frac 34 \lambda^{(1)} - C \frac{\lambda^2}{\kappa^2} - \lambda\right]
 ||u||_{L^2(\Omega)}^2 \leq 0\;.
\end{equation}
Using the assumption (\ref{x0}), this gives $u=0$ for $\lambda$ small enough and the 
proof of Theorem \ref{Theorem2.2}.

\begin{remark}\label{Remark2.6}.\\
{\rm 
Note that with a small improvement of the method, it is possible (taking $\epsilon = \frac 1\kappa$
 in (\ref{x6})  ) to show that one can choose,
in the limit $\kappa \ar + \infty$, $\lambda_0(\kappa)$ satisfying~:
 \begin{equation}\label{x8}
\lambda_0 (\kappa) \geq \lambda^{(1)} - \Og(\frac 1\kappa)\;.
\end{equation}
This will be developped in Section \ref{Section4n}.
}
\end{remark}

\begin{remark}\label{Remark2.6a}.\\
{\rm Observing that $\lambda \mapsto \frac 1 \lambda G_{\lambda,\kappa} (u,A)$ is monotonically
 decreasing, one easily obtains, that the set of $\lambda$'s such that $(0,A_e)$
 is a global minimum is an interval of the form $]0,
 \lambda_0^{opt}(\kappa)]$. Inequality (\ref{x8}) implies~:
 \begin{equation}\label{x8a}
\lambda_0^{opt}(\kappa)\geq \lambda^{(1)} - \Og(\frac 1\kappa)\;.
\end{equation}
 Similar arguments
 are used in \cite{Du} for the Abrikosov's case. We recall that, in
 this case, the domain $\Omega$, is replaced by a torus $\rz^2/\Lg$
 where $\Lg$ is the lattice
 generated over $\zz^2$ by two independent vectors of $\rz^2$.\\
Observing now that $\kappa \mapsto  G_{\lambda,\kappa} (u,A)$ is monotonically
 increasing, one easily obtains that $\kappa \mapsto
 \lambda_0^{opt}(\kappa)$ is increasing. Using (\ref{x8a}) and
 Proposition \ref{Proposition3n1}, one gets
that $\kappa \mapsto \lambda_ 0^{opt}(\kappa)$ is increasing from $0$ to
 $\lambda^{(1)}$
 for $\kappa \in ]0,+\infty[$. }
\end{remark}
\section{Estimates in  the case $\kappa$ small.}\label{Section3n}
We have already shown in Proposition \ref{Proposition2.7a} that, if
$\lambda > \lambda^{(1)}(\kappa)$, then the normal state is  not a
minimizer. In other words (see Remark \ref{Remark2.6a}), under
condition (\ref{x0}), we have~:
\begin{equation}
0 < \lambda_0^{opt} (\kappa) \leq \lambda^{(1)}\;.
\end{equation}

If we come back to the formula (\ref{x7}), one immediately obtains the
 following first result~:
\begin{proposition}\label{Proposition3n1}.\\
There exist constants $\mu_0\in]0,\lambda^{(1)}]$ and $\alpha_0 >0$
such that, for $\lambda\in]0,\mu_0]$ satisfying
\begin{equation}\label{3n1}
\lambda \leq \alpha_0 \kappa\;,
\end{equation}
the minimizer is necessarily the normal solution.
\end{proposition}

 In order to get complementary results, it is also interesting to compute the energy of the pair $(u,A)= (1, 0)$. 
This will
 give,  in some asymptotic regime, some information about the possibility for the normal solution (or later 
for a bifurcating solution)
 to correspond to a global minimum of the functional. An immediate computation gives~:
\begin{equation}\label{x10}
G_{\lambda,\kappa}(1,0) = - \frac \lambda 2 |\Omega| + 
\frac{\kappa^2}{ \lambda} \int_{\rz^2} H_e^2 dx \;.
\end{equation}
We see in particular that when $\frac \kappa \lambda$ is small, the normal solution cannot
 be a global minimizer of $G_{\lambda,\kappa}$.\\
As already observed in Subsection \ref{Subsection1.2},
what is  more relevant is probably the integral $\int_{\tilde \Omega} H_e^2\; dx$ instead of
$\int_{\rz^2} H_e^2\; dx$ in (\ref{x10}). Note also that it would be quite
 interesting to determine the minimizers in the limit $\kappa \ar 0$. 
We note indeed that $(1,0)$
 is not a solution of the GL-system, unless $H_e$ is identically zero
 in $\Omega$. 
Let us show the following proposition.
\begin{proposition}\label{Proposition3.2}.\\
If 
\begin{equation}\label{p1}
\kappa < \lambda \cdot
\left ( \frac{|\Omega|} {2 \int_\Omega H_e^2 \; dx}\right)^\frac 12\;,
\end{equation}
and if $\Omega$ is simply connected 
then the normal solution is not a global minimum.
\end{proposition}
{\bf Proof}.\\
Let $\psi_n$ a sequence of $C^\infty$ functions such that
\begin{itemize}
\item $0\leq \psi_n \leq 1$;
\item $\psi_n =0 $ in a neighborhood of  $\bar \Omega$;
\item $\psi_n (x) \ar  1, \forall x \not\in \overline{\Omega}$ ;
\end{itemize}
We observe that
\begin{equation}\label{p2}
 \int_{\rz^2} ((1-\psi_n) H_e) ^2\; dx \ar \int_\Omega H_e^2\; dx \;.
\end{equation}
We can consequently choose $n$ such that~:
\begin{equation}\label{p3}
\kappa < \lambda \cdot
\left ( \frac{|\Omega|} {2 \int_{\rz^2} \left((1-\psi_n)H_e\right)^2 \; dx}\right)^\frac 12\;,
\end{equation}
We now try to find $A_n$ such that
\begin{itemize}
\item $ \rot A_n = \psi_n H_e$;
\item $\supp A_n \cap \Omega =\emptyset$.
\end{itemize}
We have already shown how to proceed when $\Omega$ is starshaped. In the general case,
 we first choose ${\tilde A}_n$ such that~: $ \rot {\tilde A}_n = \psi_n H_e$, without
 the condition of support (see (\ref{i2}) for the argument).\\
We now observe that $\rot {\tilde A}_n = 0$ in $\Omega$. Using the simple connexity, we can find
 $\phi_n$ in $C^\infty({\bar \Omega})$ such that ${\tilde A}_n = \nabla \phi_n$. We can now extend $\phi_n$
 outside $\Omega$ as a compactly supported $C^\infty$ function in $\rz^2$ ${\tilde \phi}_n$.
We then take $A_n = {\tilde A}_n - \nabla {\tilde \phi}_n$.\\
It remains to compute the energy of the pair $(1,A_n)$ (which is strictly negative) in order to
 achieve 
 the proof of the proposition.

\begin{remark}\label{Remark3n4}.\\
{\rm In the case when $\Omega$ is not simply connected. Proposition
\ref{Proposition3.2} remains
true, if we replace $\Omega$ by ${\widetilde \Omega}$,  where $\widetilde
\Omega$  is the
smallest simply connected open set containing $\Omega$.}
\end{remark}

\begin{remark}\label{Remark3n5}.\\
{\rm It would be interesting  to see how one can use the techniques of
\cite{AfDa} for analyzing the properties of the zeros of the
minimizers, when they are not normal solutions. The link between the
two papers is given by the relation $\lambda= (\kappa d)^2 $.}
\end{remark}
In conclusion, we have obtained, the following theorem~:
\begin{theorem}.\\
Under condition (\ref{x0}), there exists $\alpha_0 >0$, such that~:
\begin{equation}
\left ( \frac{|\Omega|} {2 \int_{\widetilde \Omega} H_e^2 \;
dx}\right)^\frac 12 \leq \frac{\lambda_0^{opt}(\kappa)}{\kappa}\leq
\inf \left(\alpha_0, \frac{\lambda^{(1)}}{\kappa}\right)\;.
\end{equation}
\end{theorem}

\section{Localization of pairs with small energy, in the case $\kappa$ large.}\label{Section4n}

 When $\kappa$ is large and $\lambda -\lambda^{(1)}$ is small enough, we will
 show as in \cite{Du}
 that all the solutions of non positive energy of the GL-systems are in a 
suitable neighborhood of $(0, A_e)$ independent of
 $\kappa\geq \kappa_0>0$. This suggests
 that in this limiting regime these solutions of the GL-equations (if there exist and if they
 appear
 as local minima) will furnish global minimizers. Let us show this localization statement. The proof
 is quite similar to the proof of Theorem \ref{Theorem2.2}. We recall that we have (\ref{x2})-(\ref{x6}). Now we add the condition that, for some $\eta >0$, 
\begin{equation}\label{reg1}
\lambda \leq  \lambda^{(1)} +  \eta\;.
\end{equation}
Note that we have already solved the problem when
$\lambda \leq \lambda^{(1)} - \frac C \kappa$, so we are mainly interested in the $\lambda$'s
in an interval
 of the form  $[\lambda^{(1)} - \frac C \kappa, \lambda^{(1)} + \eta ]$.\\
The second assumption is that we consider only pairs
 $(u,A) \in H^1(\Omega) \times H^1_{loc} (\rz^2)$ such that 
\begin{equation}\label{reg1a}
G_\lambda (u, A) \leq 0\;.
\end{equation}

We improve (\ref{x6}) into
\begin{equation} \label{reg2}
||(\nabla - iA) u||_{L^2(\Omega)}^2 \geq 
\left( \left(1-\epsilon -
 \frac C\epsilon ||a||^2_{L^4(\Omega)}\right)_{_+} \,\lambda^{(1)} \;
-  \frac C\epsilon ||a||^2_{L^4(\Omega)})\right) ||u||_{L^2(\Omega)}^2\;.
\end{equation}
Taking $\epsilon = \frac 1\kappa$, we get, using also (\ref{x5a}), the existence of $\kappa_0$ and 
  $C$ such that, 
for $\lambda \in [0,\lambda^{(1)} + \eta]$  and for $\kappa\geq \kappa_0$,
\begin{equation} \label{reg3}
||(\nabla -i A) u ||^2 \geq 
\left( \left( (1- \frac{C}{\kappa}\right) \lambda^{(1)} 
-  \frac C\kappa \right) ||u||^2\;,
\end{equation}
for any $(u, A)$ such that $G_{\lambda,\kappa} (u,A)\leq 0$.\\

Coming back to (\ref{i1}), and, using again 
the negativity of the energy $G_{\lambda,\kappa} (u,A)$ of the pair $(u,A)$, we get
\begin{equation}\label{reg4}
\lambda \int_\Omega |u|^4  dx \leq (\eta +   \frac{C}{\kappa}) ||u||_{L^2(\Omega)}^2\;.
\end{equation}
But by Cauchy-Schwarz, we have
\begin{equation}\label{reg4a}
\int_\Omega |u|^2 \; dx \leq |\Omega|^\frac 12 (\int_\Omega |u|^4\; dx)^{\frac 12}\;.
\end{equation}
So we get
\begin{equation}\label{reg5}
||u||_{L^2(\Omega)} \leq (\frac {|\Omega|}{\lambda})^\frac 12 (\eta + \frac{C}{
\kappa})^\frac 12
\end{equation}
We see that this becomes small with $\eta$ and $\frac 1 \kappa$. 
It is then also easy
 to control the norm of $u$ in $H^1(\Omega)$.  We can indeed use successively (\ref{x6b}), 
 (\ref{x6a1}), (\ref{reg1a}) and the trivial inequality:
\begin{equation}
||(\nabla -i A) u||_{L^2(\Omega)}^2 \leq \lambda ||u||_{L^2(\Omega)}^2 + G_\lambda (u,A)\;.
\end{equation}
The control of $(A-A_e)$ in the suitable choice of gauge is also easy through (\ref{x2})
 and (\ref{x4a}).\\
Note also that if $\lambda<\lambda^{(1)}$, we  obtain  the better
\begin{equation}\label{reg6}
||u||_{L^2(\Omega)} \leq \frac {C}{ \kappa \lambda^\frac 12}\;.
\end{equation}
So we have shown in this section the following theorem:
\begin{theorem}\label{Theorem4.1}.\\
There exists $\eta_0 >0$  such that, for $0<\eta <\eta_0$
and  for $\lambda\leq \lambda^{(1)} + \eta$, then there exists $\kappa_0$ such that
 for  $\kappa\geq \kappa_0$,
 all the pairs $(u,A)$ with negative energy are in a suitable neighborhood 
$\Og(\eta, \frac 1\kappa)$
of the normal solution in $H^1(\Omega\;,\; \cz)\times H^1(\Omega,\rz^2)$
 whose size tends to $0$ with $\eta$ and $\frac 1 \kappa$.
\end{theorem}
\begin{remark}\label{Remark3.2}.\\
{\rm Using the same techniques as in \cite{Du}, one can also show that there
 are no solutions of the Ginzburg-Landau equations outside this neighborhood.
 This is discussed in Section \ref{Section5n}.}
\end{remark}

\section{A priori localization for solutions of Ginzburg-Landau
 equations}\label{Section5n}
In this section, we give the proof of Remarks \ref{Remark2.4} and
\ref{Remark3.2}. The proof is adapted from  Subsection 4.4 in \cite{Du} which analyzes the Abrikosov situation. Similar estimates can
 also be found in \cite{GiPh} (or in \cite{AfDa}) but in a different asymptotical
regime.

We assume that $(u,A)$ is a pair of solutions of the Ginzburg-Landau equations (\ref{i3})
 and rewrite the second Ginzburg-Landau equation, with $A= A_e + a$  in the form~:
\begin{equation}\label{b1}
L a = \frac{\lambda}{\kappa^2} \Im \left ( {\bar u}\cdot (\nabla - i (A_e +a))u\right)\;.
\end{equation}
Here $L$ is the operator defined on the space  $E^2(\Omega)$, where,
for $k\in \nz^*$,
\begin{equation}\label{b2}
E^k(\Omega):=\{ a\in H^k(\Omega;\rz^2)\;|\; \div a=0\;,\; 
 a\cdot \nu_{/\pa\Omega} = 0\;\}\;,
\end{equation}
by
\begin{equation}\label{b3}
L = \rotadj \rot = -\Delta\;.
\end{equation}
One can easily verify that $L$ is an isomorphism from $E^2(\Omega)$ onto $L^2(\Omega)$.
One first gets the following
\begin{lemma}\label{LemmaB1}.\\
If $(u,A_e +a)$ is a solution of the GL-system (\ref{i3}) for some $\lambda >0$, then
 we have~:
\begin{equation}\label{b4}
|| L a||\leq \frac {|\Omega|^\frac 12 \lambda^{\frac 32}}{\kappa^2}\;.
\end{equation}
\end{lemma}
{\bf Proof of Lemma}.\\
We start from (\ref{b1}) and using Proposition \ref{Proposition1.5}, we obtain~:
\begin{equation}\label{b5}
|| L a||^2 \leq \frac{\lambda^2}{\kappa^4} || (\nabla -i A) u||^2\;.
\end{equation}
Using the first GL-equation, we obtain~:
\begin{equation}\label{b6}
|| L a||^2 \leq \frac{\lambda^3}{\kappa^4} \int_\Omega |u|^2 (1- |u|^2) \; dx\\
\end{equation} 
Using again Proposition \ref{Proposition1.5}, we obtain the lemma.\\

So Lemma \ref{LemmaB1} shows, together with the properties of $L$, that there exists a constant $C_\Omega$
 such that
\begin{equation} \label{b7}
|| a||_{H^2(\Omega)} \leq C_\Omega \frac { \lambda^{\frac 32}}{\kappa^2}\;.
\end{equation}
This permits to control the  size of $a$ when $\lambda$ is small or $\kappa$ is large.
In particular, using Sobolev's injection Theorem, we get the existence
of a constant $C'_\Omega$ such that~:
\begin{equation} \label{b8}
|| a||_{L^\infty(\Omega)} \leq C'_\Omega \frac { \lambda^{\frac 32}}{\kappa^2}\;.
\end{equation}

The second step consists in coming back to our solution $(u,A)$ of the Ginzburg-Landau equations. Let us rewrite
 the first
 one in the form~:
\begin{equation}\label{c1}
- \Delta_{A_e} u = \lambda u (1-|u|^2) - 2 i a \cdot (\nabla - i A_e) u - |a|^2 u\;.
\end{equation}
Taking the scalar product with $u$ in $L^2(\Omega)$, we obtain:
$$
\begin{array}{ll}
\lambda \; ||\;|u|^2\;||^2 + \langle - \Delta_{A_e} u \;,\; u\rangle &\leq  \lambda ||u||^2
 + 2 ||a||_{L^\infty} ||u|| \sqrt{\langle - \Delta_{A_e} u \;,\; u\rangle} +
 ||a||_{L^\infty} ^2 ||u||^2 \\
 & \leq (\lambda + (1 + \frac 1 \epsilon) ||a||_{L^\infty} ^2) ||u||^2 +
 \epsilon \langle - \Delta_{A_e} u \;,\; u\rangle\;.
\end{array}
$$
We have finally obtained, for any $\epsilon \in]0,1[$, and any pair $(u,A)$ solution of the GL-equations
 the folllowing inequality~:
\begin{equation}\label{c2}
\lambda \int_\Omega |u(x)|^4\; dx + \langle - \Delta_{A_e} u \;,\; u\rangle \leq
 \frac{1}{1-\epsilon}\cdot (\lambda + (1 + \frac 1 \epsilon) ||a||_{L^\infty} ^2) ||u||^2 \;.
\end{equation}
Forgetting first the first term of the left hand side in (\ref{c2}), 
 we get the following alternative~:
\begin{itemize}
\item
Either $u=0$,
\item
or $$\lambda^{(1)} \leq 
 \frac{1}{1-\epsilon}\cdot (\lambda + (1 + \frac 1 \epsilon) ||a||_{L^\infty} ^2)\;. $$
\end{itemize}
If we are in the first case, we obtain immediately (see (\ref{b1}), the equation 
$La =0$ and consequently $a=0$. So we have
 obtained that $(u,A)$ is the normal solution. 

 The analysis of the occurence or not
 of the second case depends on the assumptions done in the two remarks, through (\ref{b8}) and
 for a suitable choice of
 $\epsilon$ ($\epsilon = \frac 1k$).  So we get immediately the existence of $\lambda_1(\kappa)$ and its estimate
 when $\kappa \ar + \infty$. If we now  assume (see (\ref{reg1})) that 
$\lambda \in ]\lambda^{(1)} - \eta, 
\lambda^{(1)} + \eta[$, we come back to (\ref{c2}) and write~:
$$
\lambda \int_\Omega |u(x)|^4\, dx  \leq 
\left(\frac{1}{1-\epsilon}\cdot (\lambda + 
(1 + \frac 1 \epsilon) ||a||_{L^\infty} ^2)- \lambda^{1}\right) ||u||^2 \;.
$$
Using (\ref{reg4a}), this leads to
\begin{equation}
\lambda ||u ||^2 \leq 
\left(\frac{1}{1-\epsilon}\cdot (\lambda + 
(1 + \frac 1 \epsilon) ||a||_{L^\infty} ^2)- \lambda^{(1)}\right)_+ |\Omega|\;.
\end{equation}
This shows, as in (\ref{reg5}), that $u$ is small in $L^2$ with $\eta$ and $\frac 1 \kappa$.\\
We can then conclude as in the proof of Theorem \ref{Theorem4.1}. The control of $u$ in $H^1$ is
 obtained through (\ref{c2}).

\begin{theorem}\label{Theorem5.1}.\\
There exists $\eta_0 >0$  such that, for $0<\eta <\eta_0$
and  for $\lambda\leq \lambda^{(1)} + \eta$, then there exists $\kappa_0$ such that
 for  $\kappa\geq \kappa_0$,
 all the pairs $(u,A)$ solutions of the (G-L)-equations  are in a suitable neighborhood 
$\Og(\eta, \frac 1\kappa)$
of the normal solution in $H^1(\Omega\;,\; \cz)\times H^1(\Omega,\rz^2)$
 whose size tends to $0$ with $\eta$ and $\frac 1 \kappa$.
\end{theorem}

\section{About bifurcations and stability}\label{Section6n}
\subsection{Preliminaries \label{Subsection6.1}}
Starting from one normal solution, a natural way is to see, if,  when increasing $\lambda$
 from $0$, one can bifurcate for a specific value of $\lambda$. Proposition
 \ref{Proposition2.7} shows that it is impossible before $\lambda^{(1)}$. 
A necessary condition is actually 
that $\lambda$ becomes an eigenvalue of the Neumann realization of $-\Delta_{A_e}$ in $\Omega$.
 We shall consider what is going on at $\lambda^{(1)}$.\\
Note here that there is an intrinsic degeneracy to the problem related to the existence of an $S^1$
 action. We have indeed the trivial lemma
\begin{lemma}\label{Lemma6.1}.\\
If $(u,A)$ is a solution, then $(\exp i \theta u,A)$ is a solution.
\end{lemma}
This degeneracy is independent of the gauge degeneracy.\\
In order to go further, we add the assumption
\begin{equation}\label{6.1}
\lambda^{(1)} \mbox{ is a simple eigenvalue}.
\end{equation}
In this case, we denote by $u_1$ a corresponding normalized eigenvector.\\

Now, one can try to apply the general bifurcation theory due to Crandall-Rabinovitz.
 Note that, although, the eigenvalue is assumed to be simple, it is not exactly
 a simple eigenvalue in the sense of Crandall-Rabinowitz which are working
 with real spaces. Actually, this is only simple modulo this $S^1$-action. We are not aware of
 a general theory dealing with this situation in full generality (see
 however \cite{GoSc})
 but special cases involving Schr\"odinger operators with magnetic field
 are treated in \cite{Od}, \cite{BaPhTa} and \cite{Du}. The article
 \cite{BaPhTa} is devoted to the case of the disk and \cite{Od} (more
 recently \cite{Du}) to the case of Abrikosov's states.\\
All the considered operators are (relatively to the wave function
 or order parameter) suitable realizations of operators of the type
$$u \mapsto  - \Delta_A u - \lambda  f(|u|^2) u \;,$$
with $f(0)=1$.\\

The main theorem is the following~:
\begin{theorem}\label{Theorem6.2}.\\
Under the assumptions (\ref{x0}) and (\ref{6.1}), there exist
$\epsilon_0$ and  a
bifurcating 
family of solutions $(u(\cdot;\alpha), A(\cdot;\alpha),
\lambda(\alpha))$ in $H^1(\Omega,\cz)\times E^1(\Omega)\times \rz^+$,
with $\alpha \in D(0,\epsilon_0)\subset \cz$
 for the Ginzburg-Landau equations such that
\begin{equation}
\begin{array}{rll}
u(\cdot;\alpha)& = \alpha u_1 + \alpha|\alpha|^2 u^{(3)}(\cdot;\alpha)\;,& \mbox{ with }
\langle u_1\;,\; u^{(3)}\rangle =0\;,\\
A(\cdot, \alpha)&= A_e + |\alpha|^2 a_2 + |\alpha|^4 a^{(4)}(\cdot;\alpha)\;,&\\
\lambda(\alpha)&= \lambda^{(1)} + c(\kappa) |\alpha|^2 + \Og(\alpha^4)\;.&
\end{array}
\end{equation}
Here $u^{(3)}(\cdot;\alpha)$ and $a^{(4)}(\cdot;\alpha)$ are bounded in $H^1$.\\
This solution satisfies, $\forall s\in\cz, |s|=1$~:
\begin{equation}
u(\cdot; s \,\alpha) = s \, u(\cdot;\alpha)\;,\;
A(\cdot; s  \,\alpha) = A(\cdot;\alpha)\;.
\end{equation}
Moreover, if $c(\kappa)\neq 0$,  all the solutions $(u,A,\lambda)$ of the Ginzburg-Landau equations lying in
a sufficiently small
 neighborhood in $H^1\times E^1\times \rz^+$ of $(0,
A_e,\lambda^{(1)})$ are described by
 the normal solutions $(0,A_e,\lambda)$ and the bifurcating
solutions.\\
\end{theorem}
The constant $c(\kappa)$ will be  explicited in the
next subsection.

\subsection{About the proof, construction of formal solutions.}
The starting  point is the GL-system written in the form
\begin{equation}
\begin{array}{rl}
(- \Delta_{A_e} -\lambda^{(1)} ) u &= (\lambda - \lambda^{(1)}) u - \lambda u |u|^2
 - 2 i a \cdot (\nabla - i A_e) u  - ||a||^2 u \\
L\;  a &= \frac{\lambda}{\kappa^2} \Im \left ( {\bar u}\cdot (\nabla - i A)u \right)
\end{array}
\end{equation}
We then use the standard method. We look for a solution in the form
 $$u=\alpha u_1  + \alpha |\alpha|^2 u_3 + \Og(\alpha^5)\;,$$
$$a = |\alpha|^2 \; a_2 +\Og( \alpha^4)$$
 and 
$$\lambda(\alpha) = \lambda^{(1)}
 + c (\kappa)\, |\alpha|^2 + \Og (\alpha^4) \;.$$
 We 
 can eliminate the $S^1$-degeneracy by imposing $\alpha$ real (keeping only the parity).
We refer to \cite{Du} for details and just detail the beginning of the formal proof
 which gives the main conditions.
We first  obtain, using the second equation,
\begin{equation}
a_2= \frac{\lambda^{(1)}}{\kappa^2} b_2\;,
 \end{equation}
with 
\begin{equation}
b_2:= L^{-1} \Im \left ( {\bar u_1}\cdot (\nabla - i A_e)u_1 \right)\;.
\end{equation}
Taking then the scalar product in $L^2$ with $u_1$, in the first equation, we get
 that 
\begin{equation}
c(\kappa) = \lambda^{(1)} \left( I_0 - \frac{2}{\kappa^2}  K_0 \right)  \;,
\end{equation}
with
\begin{equation}
I_0:=\int_\Omega |u_1 (x)|^4 \; dx\;,
\end{equation}
and
\begin{equation}
K_0 = - \langle  i  b_2 \cdot (\nabla - i A_e) u_1 \;,\; u_1 \rangle\;.
\end{equation}

\begin{remark}.\\
>From this expression for $c(\kappa)$, we immediately see that
 there exists $\kappa_1$ such that, for $\kappa\geq \kappa_1$,
$c(\kappa) >0$.  Moreover,   the uniqueness statement in Theorem 
\ref{Theorem6.2} is true in a
neighborhood
 which can be chosen independently of $\kappa\in [\kappa_1,+\infty[$.
\end{remark}
Let us now observe, that, $b_2$ being divergence free, it is immediate by integration by part
 that $K_0$ is real.
Computing  $\Re K_0$, we immediately obtain~:
\begin{equation}
K_0 = \Re K_0 = \langle L^{-1} J_1\;,\; J_1 \rangle\;,
\end{equation}
where $J_1$ is the current~:
\begin{equation}
J_1:= \Im \left ( {\bar u_1}\cdot (\nabla - i A_e)u_1 \right)\;.
\end{equation}

We observe that $K_0 >0$ if and only if $J_1$ is not identically $0$.
 In  the non simply connected case, we shall find a case when $J_1=0$.
 (See Lemma \ref{Lemma4.3}).\\
Following the argument of \cite{Du} (Lemme 3.4.9), let us analyze the consequences of $J_1=0$.
By assumption $u_1$ does not vanish identically. If $u_1(x_0)\neq 0$, then we can perform in a 
sufficiently small ball $B(x_0,r_0)$ centered at $x_0$, the following computation
 in polar coordinates. We write $u_1 = r(x) \exp i \theta(x)$ and get $J_1 = r(x)^2 
(A_e-\nabla \theta) = 0$.
So $A_e = \nabla \theta$ in this ball and this implies $H_e =0$ in the
 same ball. Using the properties of the zero set
 of $u_1$ in $\Omega$ \cite{ElMaQi} and the continuity of $H_e$, we then obtain $H_e = 0$ in $\Omega$.
But we know that, if $\Omega$ is simply connected, then this implies $\lambda^{(1)}=0$.
So we have the following lemma
\begin{lemma}.\\
If $\Omega$ is simply connected and $\lambda^{(1)} >0$, then $K_0 >0$.
\end{lemma}.

Coming back to the first equation and projecting on $u_1^{\perp}$, we get~:
\begin{equation}
u_3 = R_0  v_3\;,
\end{equation}
where $v_3$ is orthogonal to $u_1$ and given by~:
\begin{equation}
v_3:= 2 a_2\cdot \left((\nabla - i A_e) u_1 \right)\;,
\end{equation}
and
$R_0$ is the inverse of $(- \Delta_{A_e} -\lambda^{(1)} )$
 on the space $u_1^{\perp}$ and satisfies
$$ R_0 u_1 = 0\;.$$

We emphasize that all this construction is uniform with the parameter $\beta = \frac 1 \kappa$
 in $]0,\beta_0]$. One can actually extend analytically the equation in order to have
 a well defined problem in $[-\beta_0,\beta_0]$.
\subsection{About the energy along the bifurcating solution.}
The proof is an adaptation of \cite{Du}.  Let us just present here
 the computation of the value of the GL-functional along the bifurcating curve.
Although it is not the proof, this gives the right condition for the stability.
For this, we observe that if $(u,A_e +a)$ is a solution of the GL-system, then
 we have:
\begin{equation}
G_{\lambda,\kappa} (u,A)= - \frac \lambda 2 \int_\Omega | u| ^4 + \frac{\kappa^2}{\lambda}
 \int_\Omega |\rot a|^2 dx\;.
\end{equation}
It is then easy to get the main term of the energy of the function for $(u,A_e +a)$
 with $a (\cdot;\alpha) =|\alpha|^2 a_2(\cdot) + \Og(\alpha^4)$ and $u(\cdot;\alpha)= \alpha u_1(\cdot) + \Og(\alpha^3)$.
\begin{equation}
G_{\lambda,\kappa} (u(\cdot;\alpha),A (\cdot;\alpha))= |\alpha|^4 \left(
- \frac {\lambda^{(1)}} {2} \int_\Omega | u_1| ^4 + \frac{\kappa^2}{\lambda^{(1)}}
 \int_\Omega |\rot a_2|^2 \;dx\right) + \Og(\alpha^6)\;.
\end{equation}
Let us first analyze the structure of the term~:
\begin{equation}\label{6.16}
K_1:= \frac{\kappa^2}{\lambda^{(1)}}  \int_\Omega |\rot a_2|^2 \;dx =  \frac{\kappa^2}{\lambda^{(1)}}  
\langle L a_2, a_2\rangle\;.
\end{equation}
But we have~:
\begin{equation}\label{6.17}
K_1:= \frac{\lambda^{(1)}}{\kappa^2}
\langle L b_2, b_2\rangle = 
\frac{\lambda^{(1)}}{\kappa^2} \langle  L^{-1}\, J_1\; ,\; J_1\rangle =
\frac{\lambda^{(1)}}{\kappa^2} K_0
\;.
\end{equation}

With these expressions, we get
\begin{equation}\label{6.18}
G_{\lambda(\alpha),\kappa} (u(\cdot;\alpha),A (\cdot;\alpha))= -|
\alpha| ^4 \cdot 
\frac{\lambda^{(1)}}{2} \left(I_0 - \frac{2}{\kappa^2} K_0\right)
+ \Og(\alpha^6)\;.
\end{equation}
So we get that the energy becomes negative along the bifurcating solution for $0<|\alpha|\leq
 \rho_0$,  if the following condition
 is satisfied~:
\begin{equation}\label{sta}
\kappa^2  >  2 \frac{K_0}{I_0}\;.
\end{equation}
Another way of writing the result is~:
\begin{proposition}.\\
Under conditions (\ref{x0}) and (\ref{6.1}), then, if
\begin{equation}
\kappa^2  \neq  2 \frac{K_0}{I_0}\;,
\end{equation}
there exists $\alpha_0 >0$ such that, for all $\alpha$ satisfying $0< |\alpha|\leq \alpha_0$,
\begin{equation}
(\lambda(\alpha) - \lambda^{(1)}) G_{\lambda(\alpha),\kappa}(u(\cdot;\alpha), A(\cdot;\alpha)) < 0\;.
\end{equation}
\end{proposition}
In particular, we have shown, in conjonction with Theorem
\ref{Theorem4.1},  the following theorem~:
\begin{theorem}.\\
There exists $\eta >0$ and $\kappa_0$, such that, 
for $\kappa > \kappa_0$ and $\lambda \leq \lambda^{(1)} + \eta$,  the global minimum
 of $G_{\lambda,\kappa}$ is realized by the normal solution for $\lambda \in ]0,\lambda^{(1)}]$
 and by the bifurcating solution for $\lambda\in ]\lambda^{(1)}, \lambda^{(1)}+\eta]$.
\end{theorem}
In particular, and taking account of Remark \ref{Remark2.6a}, we have
\begin{corollary}.\\
There exists $\kappa_c$ such that the  map $\kappa \mapsto
\lambda_0^{opt}(\kappa)$ is an increasing
 function from $0$ to $\lambda^{(1)}$ for $\kappa\in [0,\kappa_c]$ and
is constant and equal to $\lambda^{(1)}$ for $\kappa\geq \kappa_c$.
\end{corollary}
\begin{remark}.\\
{\rm Note that Theorem \ref{Theorem5.1} gives an additional
information.  For $\eta$ small enough and $\kappa$ large enough, there
are actually no
 other solutions of the (GL)-equation.}
\end{remark}
\subsection{Stability}
The last point is to discuss the stability of the bifurcating solution. We expect
 that the bifurcating solution gives a local minimum of the GL-functional for $\kappa$
 large enough, and more precisely under condition (\ref{sta}). The
 relevant notion is here the notion of strict stability. Following
 \cite{BaPhTa}, we say that $(u,A)$ (with $u$ not identically $0$) is strictly stable for
 $G_{\lambda,\kappa}$ if it is a critical point, if its Hessian is
 positive and if its  kernel in $H^1\times E^{1}$  is the one
 dimensional space $\rz
 (i u\;,\;0)$.\\
We then have the following theorem~:
\begin{theorem}.\\
Under conditions (\ref{x0}),(\ref{6.1}), and if (\ref{sta}) is
satisfied, then there exists $\epsilon_0 >0$, such that, for
$0<|\alpha|\leq \epsilon_0$,
 the solution $\left(u(\cdot;\alpha), A(\cdot;\alpha)\right)$ is strictly stable.
\end{theorem}

 We refer to \cite{Du} for the detailed proof.

\section{Bifurcation from normal solutions: special case of non simply connected
 models. \label{Section7n}}
\subsection{Introduction}
In this section, we revisit the bifurcation problem in the case when
$\Omega$ is not simply connected and when the external field vanishes
inside $\Omega$. In this very particular situation which was
considered by J.~Berger and J.~Rubinstein in
\cite{BeRu} (and later in \cite{HHOO}, \cite{HHOO1}), it is
interesting to make a deeper analysis leading for example to the
description of the nodal sets of the bifurcating solution. The
situation is indeed quite different of  the results obtained by
\cite{ElMaQi}
 in a near context (but with a simply connected $\Omega$). We mainly
follow here the presentation in \cite{HHOO1} (for which we refer for
other results or points of view) but emphasize on the link
with the previous section.

\subsection{The operator $K$ \label{Subsection7.2}}
We shall now consider the specific problem introduced by 
\cite{BeRu} and consider the case
\begin{equation}\label{x20}
\supp H_e \cap {\bar \Omega}=\emptyset\;,
\end{equation}
and, in any hole $\Og_i$, the flux of $H_e$ satisfies
\begin{equation}\label{x21}
\frac{1}{2\pi} \int_{\Og_i} H_e \in \zz+\frac 12\;.
\end{equation}
Here we recall that a hole associated to $\Omega$ is a bounded connected component of 
the complementary of $\Omega$.\\

We recall in this context, what was  introduced in \cite{HHOO}. We observe that under conditions
 (\ref{x20}) and (\ref{x21}), there exists a multivalued function $\phi$ such that
 $\exp i\phi \in C^\infty(\overline{\Omega})$ and
\begin{equation}\label{a9}
d\phi = 2 \omega_A\;,
\end{equation}
where $\omega_A$ is the $1$-form naturally attached to the vector $A$.\\
We also observe  that, for  the complex conjugation
operator $\Gamma$
\begin{equation}\label{a10}
\Gamma u = {\bar u}\;,
\end{equation}
we have the general property
\begin{equation}\label{a11}
\Gamma \Delta_{A} = \Delta_{-A} \Gamma\;.
\end{equation}
Combining (\ref{a9}) and (\ref{a11}), we obtain, for the operator
\begin{equation}\label{a12}
K:= (\exp - i \phi)\; \Gamma\;,
\end{equation}
which satisfies
\begin{equation}\label{a13}
K^2 = Id\;,
\end{equation}
the following commutation relation
\begin{equation}\label{a14}
K\; \Delta_{A} = \Delta_{A} \; K\;.
\end{equation}
Let us also observe that the Neumann condition is respected by $K$. \\
As a corollary, we get
\begin{lemma}\label{Lemma4.2}.\\
If $v$ is an eigenvector of $\Delta_{A}^N$, then $K v$
 has the same property.
\end{lemma}
This shows that one can always choose an orthonormal basis of
eigenvectors $u_j$ such that $K u_j= u_j$.

\subsection{Bifurcation inside special classes.}
Following \cite{BeRu} (but inside our point of view), we look for solution of the
GL equation in the form $(u,A_e)$ with $Ku=u$. Let us observe that
\begin{equation}\label{a21}
L^2_K(\Omega;\cz):=\{ u\in L^2(\Omega;\cz)\;|\; Ku =u\}\;,
\end{equation}
is a real Hilbert subspace of $L^2(\Omega;\cz)$.\\
We denote by $H^m_K$ the corresponding Sobolev spaces~:
\begin{equation}\label{a21a}
H^m_K(\Omega;\cz) = H^m(\Omega;\cz)\cap L^2_K\;.
\end{equation}
We now observe the
\begin{lemma}\label{Lemma4.3}.\\
If $u\in H^1_K$, then $ \Im ({\bar u}\cdot (\nabla -iA_e) u) = 0$ almost everywhere.
\end{lemma}
{\bf Proof of Lemma \ref{Lemma4.3}:}\\
Let us consider a point where $u\neq 0$. Then we have $u= \rho \exp i \theta$
 with $2 \theta = \phi$ modulo $2\pi \zz$. Remembering that $A_e = \frac 12 \nabla \phi$,
 it is easy to get the property.
\qed\\
Once this lemma is proved, one immediately sees that $(u,A_e)$ (with $Ku=u$) is a solution of
 the GL system if and only if $u\in H^1_K$ and 
\begin{equation}\label{a22}
\begin{array}{l}
- \Delta_{A_e} u - \lambda u (1- |u|^2) = 0\;,\\
(\nabla - i\,A_e)u\cdot \nu = 0 \;,\; \mbox{ on } \pa \Omega\;.
\end{array}
\end{equation}
We shall call this new system the reduced GL-equation.
But now we can apply the theorem by Crandall-Rabinowitz \cite{CrRa}. By assumption (\ref{6.1}), the kernel
 of $(-\Delta_{A_e} - \lambda^{(1)})$ is now a one-dimensional real subspace in $L^2_K$. Let us denote
 by $u_1$ a normalized ``real'' eigenvector. Note that $u_1$ is unique up to $\pm 1$.
Therefore, we have the
\begin{theorem}\label{Proposition4.4}.\\
Under assumptions (\ref{6.1}), (\ref{x20}) and (\ref{x21}),
there exists a bifurcating family of solutions \break $(u(\cdot;\alpha), \lambda(\alpha)) $ in
 $H^1_K\times \rz^+$ with $\alpha\in ]-\epsilon_0,+\epsilon_0[$,  for the reduced
 GL-equation such that
 \begin{equation}\label{3.16}
\begin{array}{l}
u(\alpha) = \alpha \,u_1 + \alpha^3 \,v(\alpha)\;,\\
\langle u_1\,,\,v(\alpha)\rangle_{L^2} = 0\;,\\
||v(\alpha)||_{H^2(\Omega)} = \Og(1)\;,
\end{array}
\end{equation}
\begin{equation}\label{3.17}
\lambda(\alpha) = \lambda^{(1)} + c \alpha^2 + \Og(\alpha^4)\;,
\end{equation}
with
\begin{equation}\label{3.18}
c = \lambda^{(1)}\cdot \int_\Omega |u_1|^4\; dx\; .
\end{equation}
Moreover
\begin{equation}\label{3.19}
u(-\alpha) = - u(\alpha)\;,\; \lambda(-\alpha) = \lambda(\alpha)\;.
\end{equation}
\end{theorem}
\begin{remark}\label{Remark4.5}.\\
{\rm Note that  the property (\ref{3.19}) is what remains of the $S^1$-invariance
 when one considers only ``real'' solutions.}
\end{remark}
Let us give here the formal computations of the main terms. If we denote by $L_0$
 the operator $L_0:=- \Delta_{A_e} - \lambda^{(1)}$. Writing $v(\alpha)= u_3 +\Og(\alpha)$,
 we get~:
$$
(L_0 - c \alpha^2) (\alpha u_1 + \alpha^3 u_3) + (\lambda^{(1)}) \alpha^3 u_1|u_1|^2 = \Og(\alpha^4)\;.
$$
Projecting on $u_1$, we get (\ref{3.18}). Projecting on $u_1^{\perp}$ and denoting by $R_0$
 the operator equal to the inverse of $L_0$ on this subspace and to $0$ on $\mbox{Ker }L_0$, we get
\begin{equation}
u_3 \,= \,- \lambda^{(1)}  R_0 (u_1|u_1|^2) 
= - \lambda^{(1)}  R_0  (u_1|u_1|^2 -c u_1)\;.
\end{equation}
 
\begin{remark}\label{Remark7.5}.\\
{\rm By the uniqueness part in Theorem \ref{Theorem6.2}, we see that the solution
$(u(\cdot;\alpha),A_e)$ is actually the solution given in this
theorem.}
\end{remark}

Another remark is that
\begin{equation}\label{3.20}
G_{\lambda(\alpha),\kappa} (u(\alpha), A_e) = - \frac{\lambda^{(1)}} {2} \cdot \alpha^4 
(\int_\Omega |u_1(x)|^4 \;dx) + \Og (|\alpha|^6)\;,
\end{equation}
so that when $\alpha \neq 0$ the energy is decreasing. This is of
 course to compare with (\ref{6.16}) (note that we have $K_0=0$). Once
 we have observed this last property, the local stability of the
 bifurcated solution near the bifurcation is clear.

The second result we would like to mention concerns the nodal sets. In
the case when $\Omega$ is simply connected, the analysis of the nodal
set of $u$ when $(u,A)$ is  a minimizer of the GL-functional  is done
in \cite{ElMaQi},  using the analyticity of the solutions of
the GL-equation and techniques of Courant. 

In the non simply connected
case, very few results are known. The following theorem is true
\cite{BeRu}, \cite{HHOO1}~:
\begin{theorem}\label{Proposition4.6}.\\
Under assumptions (\ref{x5a}), (\ref{x20}) and (\ref{x21}), 
there exists $\epsilon_1 >0$ such that, for any $\alpha\in ]0,\epsilon_1]$, the nodal
 set of $u(\alpha)$ in $H^1_K$ slits $\bar \Omega$ in the sense of \cite{HHOO}. In particular, if
 there
 is only one hole, then the nodal set  of $u(\alpha)$ consists exactly in one line joining the interior
 boundary and
 the exterior boundary.
\end{theorem}

An elegant way to recover these results (see \cite{HHOO},
\cite{HHOO1}) is to lift the situation to a suitable two-fold covering
$\Omega^\Rg$.\\

{\it {\bf Acknowledgements:} We would like to thank E.~Akkermans, F.~Bethuel, C.~Bolley,
 G.~Raugel,  T.~Rivi\`ere, S.~Serfaty, M. and T. Hoffmann-Ostenhof for useful discussions.
This work is partially supported by the TMR grant FMRX-CT  96-0001 of the European Union.}

\newpage
\appendix
\section{Analysis of the various scalings.}
When considering asymptotical regimes, it is perhaps useful to have an interpretation in terms of the
 initial variables. According to the statistical interpretation of the Ginzburg-Landau
 functional (See for example \cite{BeRuSc}), the starting point is the functional
$({\tilde v},{\tilde A}) \mapsto \Fg( {\tilde v},{\tilde A})$ with~:
$$
\begin{array}{ll}
\Fg( {\tilde v},{\tilde A})&:= \frac{1}{8\pi} \int_{\rz^2} |\rot {\tilde A} - {\tilde H_e}|^2 \;
d {\tilde x}\\
& \quad + \int_{\Omega} \frac{{\hbar}^2}{4m} 
|(\nabla - i \frac{2 e}{c} {\tilde A}) {\tilde u}|^2\;
 d {\tilde x}\\
& \quad + \int_\Omega \left( a |{\tilde u} |^2 + \frac b2 | {\tilde u}|^4\right)\; d{\tilde x}\;.
\end{array}
$$
Here $a$ is a parameter which is proportional to $(T-T_c)$ (we are only interested
 in the case $a<0$) and $b$ is essentially independent of the temperature. The other
 parameters are standard~: $\hbar = \frac{h}{2\pi}$, $h$ is the Planck constant, $e$ is the charge 
 of the electron and $m$ is the mass of the electron.
With $u= \frac{b}{|a|} {\tilde u}$  and $A = \frac{2e}{c} {\tilde A}$, we obtain~:
$$
\Fg( {\tilde v},{\tilde A}) = \frac{ |a| {\hbar}^2}{4m b} \; G_{\lambda,\kappa} (u, A)\;,
$$
with 
$H_e = \frac{ 2e}{{\hbar} c} {\tilde H_e}$, $\lambda= \frac{4 m |a|}{{\hbar}^2}$
 and $\kappa=\frac{mc}{e {\hbar}}(\frac{b}{8\pi})^\frac 12$. 
Here we emphasize that between the two functionals, no change of space  variables is involved.\\

Let now compare with another standard representation of the Ginzburg-Landau functional.
We make this time the change of variables $x= \frac{\kappa}{\sqrt{\lambda}}{\hat x}$ and if
 we change $u$ and  the $1-$form corresponding to $A$ accordingly, we obtain the standard functional~:
$$
\Eg ({\hat u}, {\hat A}) = G_{\lambda,\kappa}(u,A)\;,
$$
with
$$
\begin{array}{ll}
\Eg ({\hat u}, {\hat A})& = \kappa^2 \int_{\widehat \Omega} 
\left( -|{\hat u}|^2 + \frac 12 |{\hat u}|^4\right) d{\hat x}
\\&\quad + \int_{\widehat \Omega} |(\nabla -i{\hat A}) {\hat u}|^2 d {\hat x}
\\ & \quad + \int_{\widehat \Omega} |\rot {\hat A} - {\hat H}_e |^2 d{\hat x}\;,
\end{array}
$$
with
$$
\begin{array}{ll}
{\hat H}_e &= \frac{\kappa^2}\lambda H_e\;,\\
{\widehat \Omega} & = \frac{\sqrt{\lambda}}{\kappa} \Omega\;.
\end{array}
$$
Here we observe that the open set $\Omega$ 
is not conserved in the transformation. We have to keep this
 in mind when comparing in the limit $\kappa \ar + \infty$ the contributions of Sandier and
 Serfaty \cite{SaSe} or \cite{LuPa1} with the results presented in this paper.

\end{document}